\documentclass[aps,twocolumn,amssymb,showpacs]{revtex4-1}
\usepackage{graphicx}
\usepackage{bm}
\usepackage{amsmath}
\usepackage{amssymb}
\usepackage{lipsum}

\def\be{\begin{equation}}
\def\en{\end{equation}}

\def\ve{\varepsilon} 
\newcommand{\bi}[1]{\mbox{\boldmath$#1$}}

\def\p{\partial}
\def\bea{\begin{eqnarray}}
\def\ena{\end{eqnarray}}

\begin{document}

\title{Nano-domain formation in charged membranes: Beyond Debye-H\"{u}ckel approximation} 

\author{Ryuichi Okamoto$^{1}$}
\author{Naofumi Shimokawa$^{2}$}  
\author{Shigeyuki Komura$^{1}$}

\affiliation{
$^{1}$ Department of Chemistry, Graduate School of Science and Engineering,
Tokyo Metropolitan University, Tokyo 192-0397, Japan \\
$^{2}$ School of Materials Science, Japan Advanced Institute of Science and Technology, 
Ishikawa 923-1292, Japan}

\begin{abstract}
We investigate the microphase separation in a membrane composed of charged lipid, 
by taking into account explicitly the electrostatic potential and the ion densities in the 
surrounding solvent.  
While the overall (membrane and solvent) charge neutrality is assumed, the membrane can have a 
non-zero net charge. 
The static structure factor in the homogeneous state is analytically obtained without using the Debye-H\"uckel approximation and is found to have a peak at an intermediate wave number. 
For a binary membrane composed of anionic and neutral lipids, the characteristic wave number corresponds 
to a scale from several to tens of nanometers. 
Our numerical calculation further predicts the existence of nano-domains in charged membranes.
\end{abstract}

\pacs{87.16.D-, 82.45.Gj, 87.16.dt}

\maketitle

\section{Introduction}

Much attention has been paid to phase separations in artificial multi-component lipid membranes. 
In these systems, many degrees of freedom such as lipid composition and
membrane shape deformation are coupled to each other, 
leading to complex phase behaviors~\cite{Komura14}.
In particular, long-lived small domains in membranes may play important biological 
roles~\cite{Simons97,Simons10,Arumugam15}.
In general, charge-induced microphase formation has been intensively investigated in 
soft matter such as polyelectrolytes~\cite{Borue,Joanny}, electrolyte fluid 
mixtures~\cite{Sadakane,Onuki1,Onuki2}, and charged Langmuir monolayers~\cite{Kato, Wilke}.
Several authors have studied the microphase 
formation in  membranes composed of both anionic and cationic lipids for which the 
net charge within the membrane vanishes~\cite{Solis, Velichko, Safran}.

In the last decade, researchers have investigated phase separations in giant unilamellar 
vesicles (GUVs) composed of anionic and neutral lipids~\cite{ShimokawaEx,Dimova1,Dimova2,Keller,Himeno}. 
In these experiments, GUVs have a \textit{non-zero} net charge. 
From a theoretical viewpoint, Guttman and Andelman originally predicted  a \textit{microphase} 
separation in binary charged membranes within the Debye-H\"{u}ckel (DH) approximation~\cite{Guttman}. 
However, no evidence for a charge-induced microphase formation 
on the scale of the optical resolution was found in refs.~\cite{ShimokawaEx,Dimova1,Dimova2,Keller,Himeno}.
Hence some theories for a \textit{macrophase} separation in such charged membranes have been  
developed~\cite{May1,May2,Shimokawa}. 
Recently,  Puff \textit{et al.} reported the formation of nanoscale domains,
whose scale is smaller than the optical resolution,  with the addition of ganglioside GM1~\cite{Puff}. 
In their mixtures, GM1 is anionic while the other components are all neutral.
Given these experimental observations, it is necessary to study the competition between the macrophase 
and microphase separations in charged membranes, as well as the characteristic length scale 
associated with the microphase separation.

The DH approximation is justified when $b\kappa \gg 1$, where $b$ and $\kappa$ 
are the Gouy-Chapman length and the Debye wave number, respectively. 
Notice that $b$ is inversely proportional to the surface charge density. 
In a strongly segregating charged membrane, each domain usually has a large 
surface charge density, and thus the DH approximation is no more valid~\cite{Safran}.
Furthermore, for membranes that have non-zero net charge, the DH approximation is 
inapplicable even to a disordered phase nor to weakly segregating domains. 
For a binary membrane composed of anionic and neutral lipids (as discussed later in more
detail), we can estimate $\phi b\sim 1$~\AA\ with $\phi$ being the fraction of the anionic lipid, whereas for a 1:1 electrolyte solution, $\kappa$ is in the range $10^{-3}$--$10^{-1}$~\AA$^{-1}$. 
Hence the DH condition $b\kappa\gg1$ is not satisfied unless $\phi$ is very small,  and it is imperative 
to go beyond it.

In this Letter, we investigate the microphase formation in binary membranes composed of 
charged lipids for general ionic strength. 
We assume the overall (membrane and solvent) charge neutrality, while the membrane can 
have a non-zero net charge. 
Solving the full non-linear Poisson-Boltzmann equation (PBE), we discuss (i) the scale of  the 
microphase structures, and (ii) the conditions for the microphase formation. 
Our theory predicts a {\it microphase endpoint} (MEP) \cite{Olmsted} in the composition-temperature plane at which an end of the macrophase spinodal line meets that of the microphase spinodal line. This point cannot be obtained within the DH approximation.
We also find that the characteristic length scale of the microphase separation is in the range from several 
to tens of nanometers except in the vicinity of the MEP.

\section{Free energy and PBE}
 
\begin{figure}
\includegraphics[scale=0.44]{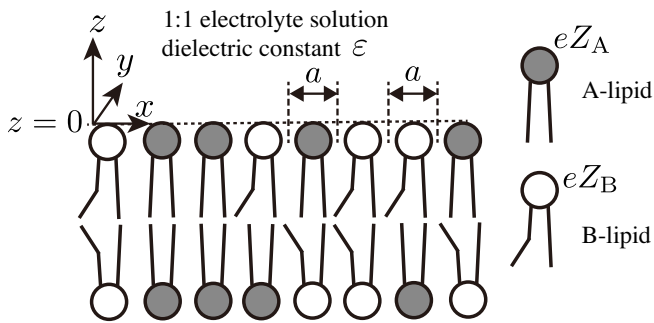}
\caption{Schematic illustration of a cross section of a charged membrane in an electrolyte solution of the dielectric constant $\ve$. 
The electric charges of A-lipid and B-lipid are $eZ_{\rm A}$ and $eZ_{\rm B}$, where $Z_{\rm J}$ 
(${\rm J}={\rm A},{\rm B}$) is the valence number and $e$ is the elementary charge. 
Both lipid species are assumed to have a common size $a$ of the hydrophilic head group.}
\label{Figmem}
\end{figure}

As shown in fig.~\ref{Figmem}, we consider a flat fluid membrane composed of A- and B-lipid 
molecules having electric charges $eZ_{\rm J}$ $({\rm J}={\rm A, B})$, where $e$ is the 
elementary charge and $Z_{\rm J}$ is the valence number.
In water, these lipid molecules form a bilayer structure.
Here the hydrophobic tails face each other and the hydrophilic head groups are in contact with water. 
Any interactions between different monolayers are neglected although there are situations 
in which inter-monolayer coupling plays a role~\cite{May2,Shimokawa}. 
Then we are allowed to consider only a two-dimensional (2D) monolayer located at $z=0$  
which is in contact with the solvent occupying the region of $z>0$ in a three-dimensional (3D) space. 
We use the abbreviations ${\bm x}=(x,y)$ and ${\bm X}=(x,y,z)$.

We assume that the molecular area $a^2$ occupied by an A-lipid is the same as that occupied by a B-lipid. 
The area fraction of A-lipid is denoted by $\phi({\bm x})$, and  that of B-lipid is  
$1-\phi({\bm x})$ under the areal incompressibility condition.  
For simplicity, we assume that the solvent is a 1:1 dilute electrolyte solution. 
The cation and anion number densities are denoted by $n_+({\bm X})$ and $n_-({\bm X})$, 
respectively. 
In our work, the overall charge neutrality is assumed;  
$\int{\rm d}^2x\, [Z_{\rm A}\phi+Z_{\rm B}(1-\phi)]+\int{\rm d}^3X\, (n_+-n_-)=0$,
where the integration $\int {\rm d}^2x $ is taken over the range $-\infty<x,y<\infty$,
while $\int {\rm d}^3X$ is for the range $z>0$ and $-\infty <x, y<\infty$.
The total free energy functional $F=F_{\rm m}+F_{\rm b}$ is given by the sum of the membrane contribution,
$F_{\rm m}$ (in the absence of the electrostatics contribution), and the bulk contribution, $F_{\rm b}$. 
The former is given by
\begin{align}
\frac{F_{\rm m}}{T}=\int {\rm d}^2x \, \left[ f(\phi)+\frac{c}{2}(\nabla_\parallel \phi)^2 \right], 
\label{Fl}
\end{align}
where $T$ is the temperature (we have set the Boltzmann constant $k_{\rm B}$ to unity), 
$f(\phi)$ is the scaled free energy (without the electrostatic contribution) per unit area 
for a homogeneous state, and the second term is the standard square-gradient form with a 
positive dimensionless coefficient $c$.
Note that $\nabla_\parallel=(\p_x, \p_y)$ is the 2D gradient operator.

The charged lipids in the membrane and the ions in the solvent generate a gradient of the electrostatic potential 
$\Psi({\bi X})$ satisfying the Poisson equation,
$\varepsilon \nabla^2 \Psi=-4\pi e(n_+-n_-)$, where $\varepsilon$ is the dielectric constant of 
water, and $\nabla=(\p_x, \p_y, \p_z)$ is the 3D gradient operator. 
The areal charge density of the membrane is given by 
$\sigma(\phi)=[Z_{\rm B}+(Z_{\rm A}-Z_{\rm B})\phi]/a^2$ in units of $e$. 
Then the boundary condition at $z=0$ is $\varepsilon \p_z \Psi|_{z=0}=-4\pi e \sigma(\phi)$. 
The bulk part of the free energy $F_{\rm b}$ consists of the entropy of ions and the electrostatic 
energy,
\begin{align}
\frac{F_{\rm b}}{T}=\int {\rm d}^3 X \, \left[ \sum_{i=\pm}n_i\{ \ln(n_i\lambda_i^3)-1\}+
\frac{\varepsilon}{8\pi T}(\nabla \Psi)^2 \right],
\end{align}
where $\lambda_i$ is the thermal de Broglie length of the $i$-th ion species.

In the mean field theory, the equilibrium state is obtained by minimizing $F$ with respect to 
$n_i$ and $\phi$ under the constraints that all the density variables are conserved quantities.  
With the aid of the relation 
$\delta (\varepsilon |\nabla \Psi|^2)=2\varepsilon \nabla\cdot (\Psi \nabla \delta\Psi)+
8\pi e \Psi (\delta n_+-\delta n_-)$, 
we can derive the nonlinear PBE as the minimization condition with respect to $n_i$~\cite{Israelachivili}, 
\begin{align}
\nabla^2 \psi=\kappa^2 \sinh \psi, 
\label{PBE}
\end{align}
where $\psi=e\Psi/T$ is the dimensionless potential, 
$\kappa=[8\pi e^2 n_\infty /(\varepsilon T)]^{1/2}$ is the Debye wave number with 
$n_\infty$ being the ion density far from the membrane ($z\to\infty$). 
Without loss of generality, we can impose the boundary condition $\psi \to 0$ as $z \to \infty$. 
Minimization with respect $\phi$ yields
\begin{align}
\frac{h}{T} \equiv \frac{\partial f}{\partial \phi}-c\nabla_\parallel^2\phi+
\psi \frac{\partial \sigma}{\partial \phi}={\rm const.}
\label{LipidEQ} 
\end{align}
In equilibrium, the grand potential 
$\Omega=F-\int{\rm d}^2x\, \sum_i n_i \mu_{i\infty}-h\int{\rm d}^3X\, (\phi-\phi_0)$ 
should be minimized. 
Here $\mu_{i\infty}=T\ln (n_\infty \lambda_i^3)$ is the ion chemical potential far from the 
membrane and $\phi_0$ is the average composition.

\section{Fluctuations around the homogeneous state} 

At high temperatures, the translational entropy of the lipid molecules dominates inter-molecular 
interactions, leading to a homogeneous phase, $\phi=\phi_0$. 
In this state, all the variables are constant in the lateral $xy$-direction, while $n_i$  
and $\psi$ depend on $z$ because of the electric charges of the lipid molecules. 
Therefore $\psi$ obeys a one-dimensional (1D) PBE $\partial_z^2 \psi=\kappa^2\sinh \psi$ with the 
boundary condition $\partial_z \psi |_{z=0}=-4\pi\ell\sigma_0$, where  
$\sigma_0=\sigma(\phi_0)$ and $\ell=e^2/\varepsilon T$ is the Bjerrum length. 
It is convenient to introduce a dimensionless number $\eta = \kappa/(2\pi\ell\sigma_0)$
which can be either positive or negative. 
Note here that the Gouy-Chapman length is given by $b =1/(2\pi \ell |\sigma_0|)=|\eta|/\kappa$.

The 1D PBE has a well-known exact solution \cite{Israelachivili},
\begin{align}
\psi_0(z)=2\ln \frac{1+\Gamma e^{-\kappa z}}{1-\Gamma e^{-\kappa z}}. 
\label{PB1D}
\end{align}
In the above, the dimensionless number $\Gamma$ is the root of $\Gamma^2+2\eta \Gamma-1=0$ and is given by
$\Gamma=-\eta \pm \sqrt{\eta^2+1}$ $(\eta \gtrless 0)$. 
The ion densities in equilibrium are expressed as $n_{\pm0}(z)=n_\infty e^{\mp \psi_0(z)}$.
In the DH condition $|\eta|\gg1$ which corresponds to a high salt and/or small surface charge condition, 
we have $\Gamma \simeq 1/(2\eta)$ and obtain $\psi_0\simeq 2e^{-\kappa z}/\eta$.

In order to see the fluctuations around the homogeneous state, we superimpose the variations, 
$\phi_0\to \phi_0+\delta\phi({\bm x})$ and $n_{i0}(z) \to n_{i0}(z)+\delta n_i({\bm X})$. 
We then examine the free energy deviation $\Delta F$ up to the bilinear order in the variations, 
$\delta\phi$ and $\delta n_i$. 
Since we are interested in the fluctuations of $\delta\phi$, we further minimize $\Delta F$ with 
respect to $\delta n_i$. 
Here we introduce the in-plane Fourier transform of a function $g({\bi x})$ as 
$g_{\bm k}=\int  {\rm d}^2 x\,  e^{-i{\bm k}\cdot {\bm x}} g({\bi x})$, 
where ${\bm k}=(k_x,k_y)$ is the 2D wave vector. In the Fourier space, $\Delta F$ is written as
\begin{align}
\frac{\Delta F}{T}=\frac{1}{2}\int \frac{{\rm d}^2k}{(2\pi)^2}  
\left[ \frac{\p^2 f(\phi_0)}{\p\phi_0^2}+ck^2 + \theta_0^2 P_k(0)\right] |\delta\phi_{\bm k}|^2, 
\label{Fbilinear}
\end{align}
where $k=|{\bm k}|$ and $\theta_0=\partial \sigma(\phi_0)/\partial \phi_0=(Z_{\rm A}-Z_{\rm B})/a^2$. 
In the above, $P_k(0)=P_k(z=0)$ where $P_k(z)$ satisfies 
\begin{align}
\left[\p_z^2-k^2-\kappa^2\cosh \psi_0(z)\right]P_k(z)=0, \label{Pk}
\end{align}
with the boundary condition $\partial_z P_k(z)|_{z=0}=-4\pi\ell$.
One can easily find that 
$\psi_0+\theta_0 \int {\rm d}^2k/(2\pi)^2 \, e^{i{\bm k}\cdot{\bm x}}P_k(z) \, \delta\phi_{\bm k}$ 
is a solution of the PBE under the boundary condition 
$\partial_z\psi(z)|_{z=0}=-4\pi\ell[\sigma_0+\theta_0\delta \phi({\bm x})]$ when the surface charge 
heterogeneity $\theta_0\,\delta\phi({\bm x})$ is sufficiently small.

In the DH condition $|\eta|\gg 1$ , we can set $\cosh \psi_0\simeq1$ in eq.~(\ref{Pk}) and obtain
\begin{align}
P_k(z)&\simeq \frac{4\pi \ell}{\sqrt{k^2+\kappa^2}} 
\exp \left(-z\sqrt{k^2+\kappa^2}\right). \label{PkDH}
\end{align}
The above expression was obtained in ref.~\cite{Guttman} and used in 
simulations~\cite{Velichko}. 
Notice that $\theta_0^2 P_k(0)$ in eq.~(\ref{Fbilinear}) does not depend on $\phi_0$ within the DH condition.

\section{Perturbation solution of nonlinear PBE}

In order to go beyond the DH approximation and discuss general values of $\eta$, we may seek the 
solution of eq.~(\ref{Pk}) perturbatively in powers of $k^2$.
This is because the solution is $P_0(z)=\partial \psi_0/\partial \sigma_0$ for $k=0$.
Introducing the dimensionless quantities $\bar z=\kappa z$ and $\bar k=k/\kappa$, we substitute 
$P_k(z)=R(\bar z)P_0( z)$ into eq.~(\ref{Pk}) and obtain 
\begin{align}
R''+(\ln P_0^2)' R' -\hat{\epsilon} \bar k^2R=0, 
\label{Req}
\end{align}
where the prime denotes the derivative with respect to $\bar z$, and we have introduced the 
``book keeping parameter" $\hat{\epsilon}$ which will be set to unity at the end. 
We expand $R$ in powers of 
$\hat{\epsilon}$ such that $R=R_0+\hat{\epsilon} R_1+\cdots$. 
Then up to the first order in  $\hat{\epsilon}$, we have
\begin{align}
&R_0''+(\ln P_0^2)'R_0'=0 \label{R0},\\
&R_1''+(\ln P_0^2)'R_1'=\bar k^2R_0, 
\label{R1}
\end{align}
and we may obtain the perturbation solution. 
However, it turns out that a secular term appears in $\hat{\epsilon} R_1$, and hence the perturbation 
solution is only \textit{locally} valid in the vicinity of the boundary, but not \textit{uniformly} valid in 
the entire region $z>0$.

In order to cure the breakdown of such a perturbation calculation,  apparently different but almost 
equivalent ways have been developed~\cite{Bender,Chen2,Kunihiro1, Kunihiro2}. 
Among these we use the renormalization group (RG) method in refs.~\cite{Kunihiro1, Kunihiro2}. 
Given the solution $G(\bar z_0)$ at any point 
$\bar z_0>0$, we first seek the solution of eqs.~(\ref{R0}) and (\ref{R1}) such that 
$P_k=RP_0\to0$ as $\bar z\to\infty$. 
Some calculation yields 
\begin{align}
&R(\bar z;\bar z_0)\nonumber \\
&=G\left[ 1-\frac{\hat{\epsilon}\bar k^2}{2} \left\{ (\bar z-\bar z_0)+
\frac{\Gamma^2}{2}(e^{-2\bar z}-e^{-2\bar z_0})\right\}\right]+O(\hat{\epsilon}^2). 
\label{Perturb1}
\end{align}
Here the term proportional to $\hat{\epsilon}(\bar z-\bar z_0)$ is a secular term that becomes 
large as $\bar z-\bar z_0$ is increased, leading to the breakdown of the regular perturbation calculation. 
To obtain a uniformly valid solution, we impose in eq.~(\ref{Perturb1}) the RG equation, 
$\partial R/\partial \bar z_0|_{\bar z=\bar z_0}=0$.
This yields the differential equation for $G$ as
\begin{align}
\frac{\partial G}{\partial\bar z_0}+\frac{\hat{\epsilon}\bar k^2}{2}
\left(\Gamma^2e^{-2\bar z_0}-1\right)G=0. 
\label{RGE}
\end{align}
Using the solution $G(\bar z_0)$, we obtain the improved solution as 
$R(\bar z;\bar z_0=\bar z)=G(\bar z)$, which is uniformly valid up to the order of 
$\hat{\epsilon}$~\cite{Kunihiro2}. 
Geometrically, the improved solution $G(\bar z)$ is the envelope of the family 
of curves $\{R(\bar z;\bar z_0)\}_{\bar z_0}$ parametrized by $\bar z_0$. 
It is tangent at each point $\bar z=\bar z_0$ to a member of the family $R(\bar z;\bar z_0)$ 
that is locally valid in the vicinity of $\bar z_0$~\cite{Kunihiro1}.

Solving eq.~(\ref{RGE}) under the boundary condition $\partial P_k/\partial\bar z|_{\bar z=0}=-2/(\eta\sigma_0)$, we finally obtain $P_k(z)$ as 
\begin{align}
P_k(z)\simeq \frac{P_0(z) }{1+\gamma\bar k^2} \exp \left[-\frac{\bar k^2}{2} \left\{ \bar z+
\frac{\Gamma^2}{2}\left(e^{-2\bar z}-1\right) \right\}\right] ,
\label{PkApp}
\end{align}
where $\gamma= \eta^2\Gamma/(\eta+\Gamma)$. 
For the DH condition $|\eta|\gg 1$, eq.~(\ref{PkApp}) reduces to eq.~(\ref{PkDH})
within the approximation $(k^2+\kappa^2)^{1/2}\simeq\kappa(1+\bar k^2/2)$. 
For a general value of $\eta$, it can be shown that eq.~(\ref{PkApp}) provides a good approximation for any $z>0$ if $\bar k \ll1$. Even if $\bar k \ll1$ is not satisfied, it is still a good approximation for sufficiently small $z$ satisfying $\bar{k}^2(\eta \Gamma+\Gamma^2\bar{z})^2 \ll1$. This region, where eq.~(\ref{PkApp}) is accurate for not so small $\bar k$, indeed exists if ($\bar{k}\eta\Gamma)^2\ll1$. For a low salt and/or large surface charge condition, $|\eta|\ll1$, the region is given by $k^2(b+z)^2\ll1$ if $(kb)^2\ll1$. 
Equation~(\ref{PkApp}) is enough for our purpose because we only need $P_k(0)$ to calculate 
the compositional structure factor (see eq.~(\ref{Fbilinear})).

\section{Conditions for microphase separation}

Using the definition $p =P_0(z=0)=2/[\sigma_0(\Gamma+\eta)]$, we substitute 
$P_k(0)\simeq p/(1+\gamma \bar k^2)$ into eq.~(\ref{Fbilinear}) and obtain 
the structure factor $S(k)=\langle |\delta\phi_{\bm k}|^2\rangle$. 
Here both $p$ and $\gamma$ are positive. 
The positivity of $p$ means that the electrostatic interaction tends to prevent the instability 
towards a macrophase separation~\cite{Shimokawa}. 
In addition, the positivity of $\gamma$ implies the possibility of the microphase formation 
when the temperature is decreased. 
If $c<\theta_0^2 p \gamma/\kappa^2$ is satisfied, $S(k)$ takes a 
maximum value at an intermediate wave number
\begin{align}
k^{\ast}=\frac{\kappa}{\gamma^{1/2}}\left[ \left(\frac{\theta_0^2 p \gamma}
{c \kappa^2}\right)^{1/2}-1\right]^{1/2}. 
\label{kmeso}
\end{align}
When the temperature is decreased to a certain value, $S(k^*)$ diverges and the modes 
$\delta\phi_{{\bm k}}$ with $|{\bm k}|=k^*$ become unstable. 
This leads to the microphase separation characterized by a typical wave length $2\pi/k^*$.

For the sake of further discussion, we assume that the free energy density $f(\phi)$ is 
given by the Bragg-Williams form, $f(\phi)=[\phi\ln\phi+(1-\phi)\ln(1-\phi)+\chi\phi(1-\phi)]/a^2$, 
where $\chi$ is the dimensionless interaction parameter that is roughly proportional to $1/T$. 
With this choice, the instability condition towards the microphase separation is written as 
$\chi >\chi^{\ast}$ with
\begin{align}
\chi^{\ast}=\chi_{\rm s}
-\frac{a^2 c \gamma}{2\kappa^2}(k^*)^4. 
\label{chiS}
\end{align}
Here $\chi_{\rm s}=[\{\phi_0(1-\phi_0)\}^{-1}+a^2\theta_0^2p]/2$ is the spinodal for the macrophase 
separation when $\theta_0^2 p \gamma/\kappa^2<c$. Equation (\ref{chiS}) defines the {\it microphase endpoint} (MEP) \cite{Olmsted}, $\chi_{\rm E}=\chi^*=\chi_{\rm s}$, at which an end of the macrophase spinodal meets that of the microphase spinodal (MEP can be a Lifshitz point when it is also the critical point of the macrophase separation).

\section{Membranes composed of anionic and neutral lipids}

\begin{figure}
\includegraphics[scale=0.45]{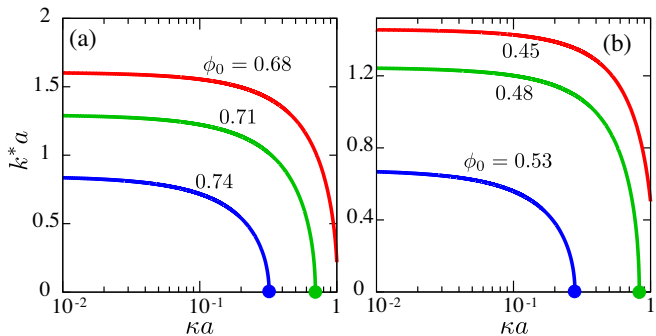}
\caption{The scaled characteristic wave number $k^* a$ as a function of the scaled Debye wave 
number $\kappa a$ for (a) $c=0.15$ and  (b) $0.4$. 
The filled circles are the microphase endpoints (MEPs) at which $k^* a$ vanishes, and correspond to $\kappa_{\rm E}$.
}
\label{Figkmeso}
\end{figure}

Following the previous experiments~\cite{ShimokawaEx,Dimova1,Dimova2,Keller,Himeno, Puff},
we hereafter set $Z_{\rm A}=-1$ and $Z_{\rm B}=0$, i.e.,
each A-lipid has a negative charge $-e$ while the B-lipid is neutral (cationic lipids are not so common in biomembranes). 
For typical lipid membranes in water, we set $\ell=7$~\AA\ and 
$a=8$~\AA~\cite{Kucerka11,Petrache04}.
Then we have $\phi_0 b=a^2/(2\pi\ell)=1.46$~\AA. 
The Debye wave number $\kappa$ is usually in the range $
10^{-3}$--$10^{-1}$~\AA$^{-1}$ for 1:1 electrolyte solutions. 
The parameter $\eta <0$ is then estimated to be $\phi_0\vert \eta \vert \sim 10^{-3}$--$10^{-1}$. 
Hence the DH condition $|\eta|\gg1$ is not satisfied even for a high salt solution, except for a very 
small fraction $\phi_0$ of the charged lipid. 
In neutral membranes, the line tension 
$\tau \sim T(c/a^2)^{1/2} $ between the coexisting phases has been measured to be 
several pN~\cite{Tian07}.
Then we may estimate that $c$ is in the range $0.1$--$1$.

In the present case, we can rewrite eq.~(\ref{kmeso}) as
\begin{align}
k^*=\frac{\kappa}{\gamma^{1/2}}\left[ \left\{ \frac{|\Gamma|(\phi^*/\phi_0)^3}{(\Gamma+\eta)^2} \right\}^{1/2}-1\right]^{1/2} \label{kmeso2}
\end{align}
with $\phi^{\ast} = [a^2/(2c\pi^2\ell^2)]^{1/3}$. 
Here the factor $|\Gamma|/(\Gamma+\eta)^2$ is a monotonic decreasing function of $|\eta|$ and is less 
than unity. 
Therefore, $k^*$ cannot exist for any $\kappa$ when $\phi_0>\phi^*$.
In fig.~\ref{Figkmeso}, the maximum wave number $k^*$ is plotted as a function of $\kappa$, 
where we set (a) $c=0.15$ (corresponding to $\phi^{\ast}=0.76$) 
and (b) $c=0.4$ ($\phi^{\ast}=0.4$).  
For $\kappa b \ll1$, we may set $\eta\simeq 0$, $\Gamma \simeq -1$ and $\gamma \simeq \eta^2$ 
in eq.~(\ref{kmeso2}) to obtain $k^* b\simeq [(\phi^{\ast}/\phi_0)^{3/2}-1]^{1/2}$, which is independent of 
$\kappa$. 
As $\kappa$ is increased, $k^*$ decreases and eventually vanishes at 
$\kappa=\kappa_{\rm E}(\phi_0)$ (marked with filled circles in fig.~\ref{Figkmeso})
when the MEP condition $(\phi^{\ast}/\phi_0)^3=(\Gamma+\eta)^2/|\Gamma|$ holds.
It should be noted that the predicted value of $2\pi/k^*$ corresponds to the scale in the range from 
several to tens of nanometers unless $\kappa$ is very close to $\kappa_{\rm E}$.

\begin{figure}
\includegraphics[scale=0.45]{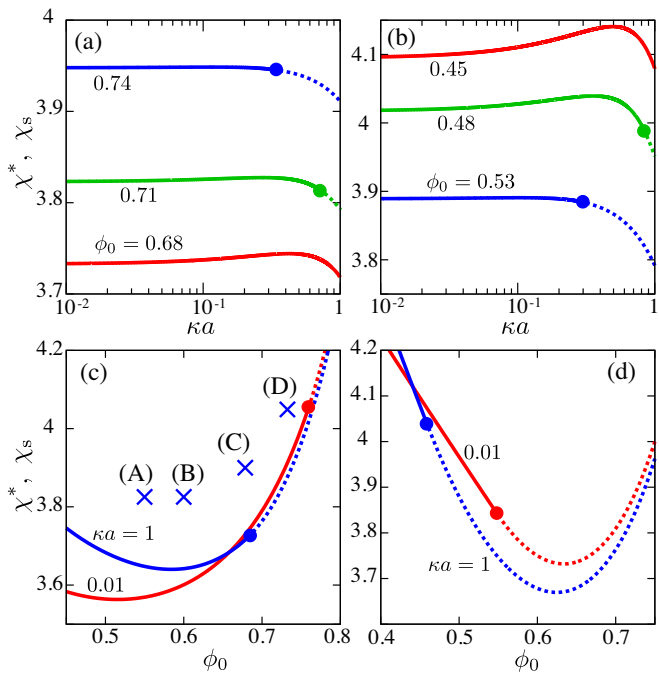}
\caption{Stability diagrams in the $(\kappa$,$\chi)$-plane ((a) and (b)) and in 
the $(\phi_0$,$\chi)$-plane ((c) and (d)). 
The parameters are $c=0.15$ in (a) and (c), and $c=0.4$ in (b) and (d). 
The spinodal lines of the microphase separation are plotted with solid lines, 
while those of the macrophase separation are shown by the dotted lines. 
The MEP for each curve is marked with a filled circle.
The four cross marks in (c) correspond to the numerical simulations in fig.~\ref{Figsim}.}
\label{Figchis}
\end{figure}

In figs.~\ref{Figchis}(a) and (b), we plot $\chi^{\ast}$ as a function of $\kappa$ for 
$c=0.15$ and $0.4$, respectively.  
All the curves for different $\phi_0$ values exhibit non-monotonic dependence on $\kappa$ 
(though it is not apparent for $\phi_0=0.74$ in (a) and for $\phi_0=0.53$ in (b)). 
When $\kappa$ is varied for fixed $\phi_0$ values, one can show that $\chi^{\ast}$ takes 
a maximum value when $\kappa$ satisfies $c=2b^2\theta_0^2 \Gamma^3/\sigma_0$ if the 
inequality $c<2b^2\theta_0^2/|\sigma_0|$ holds.
In these plots,  the MEP is located at $\chi_{\rm E}=\chi^*(\kappa_{\rm E})$
(marked with filled circles).

In figs.~\ref{Figchis}(c) and (d), we plot $\chi^{\ast}$ as a function of $\phi_0$ for 
$c=0.15$ and $0.4$, respectively. 
For $\kappa a=0.01$ and for $\phi_0\gtrsim 0.01$, we are allowed to set $\Gamma\simeq-1$ 
and $\eta\simeq0$ in eq.~(\ref{kmeso2}). 
In this case, the MEP is almost located at  
$(\phi_0, \chi)_{\rm E}=(\phi^{\ast}, \chi^{\ast}(\phi^{\ast}))$.
For a larger value of $\kappa$, $\phi_{\rm E}$ exhibits a downward shift. 
Since $k^*$ does not depend on $\phi_0$ within the DH approximation, it cannot  predict
any MEP on the ($\phi_0$, $\chi$)-plane (see also below eq.~(\ref{PkDH})). 
Indeed, within the DH approximation, the condition for the existence of $k^*$ is given by 
$c<2\pi\ell \theta_0^2/\kappa^3$ that is independent of $\phi_0$. 
With our choice of $\ell$ and $a$, this inequality becomes $\kappa a <1.76c^{-1/3}$ and is 
always satisfied for the parameter values used in figs.~\ref{Figkmeso} and \ref{Figchis}.  
In general, the DH approximation overestimates the possibility 
of the microphase separation.

\section{Numerical simulation}

Finally, we numerically integrate the equilibrium conditions, eqs.~(\ref{PBE}) and (\ref{LipidEQ}). 
We set  $Z_{\rm A}=-1$ and $Z_{\rm B}=0$ as before.
We prepare in the $xy$-plane a membrane of size $(85.3a)^2$ that is in contact with an 
electrolyte solution at $z=0$.  
The size $L_z$ of the solvent container is set to $16.25 a$. 
At $z=L_z$, we impose the boundary condition, $\psi(L_z)=0$,  which is justified 
when $\kappa L_z\gg1$. 
A periodic boundary condition is employed in the lateral $xy$-directions. 
We then solve simultaneously the fictitious dynamic equations, 
$\partial \phi/\partial t=-h+\langle h\rangle$ and 
$\partial \psi/\partial t =\nabla^2 \psi-\kappa^2\sinh \psi$, where 
$\langle \cdots\rangle=\int {\rm d}^2x \, (\cdots) /\int {\rm d}^2x$ 
denotes the areal average in the membrane. 
Although the dynamics itself has no physical meaning, the equilibrium pattern can be efficiently 
obtained as a stationary state.

\begin{figure}
\includegraphics[scale=0.42]{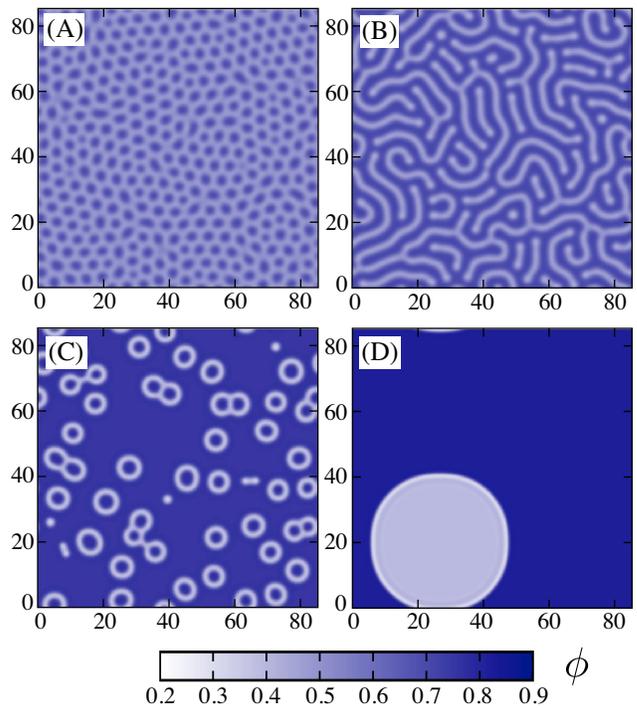}
\caption{Equilibrium composition profiles $\phi(\bm x)$. 
The membrane size is measured in units of $a$. 
The values of $\chi$ and $\phi_0$ are (A) $(\phi_0,\chi)=(0.55, 3.83)$, 
(B) $(0.6, 3.83)$, (C) $(0.68, 3.9)$ and (D) $(0.73, 4.05)$ while  $c=0.15$ and 
$\kappa a = 1.0$ are fixed. 
These four cases are marked with crosses (A)--(D) in fig.~\ref{Figchis}(c). 
Corresponding to these cases, we have (A) $(k^*a,\chi^{\ast})=(1.90,\ 3.65)$,  
(B) $(1.65,\ 3.64)$ and (C) $(0.219,\ 3.72)$, whereas in (D) the characteristic 
wave number $k^*$ does not exist and the macrophase separation occurs at $\chi_{\rm s}=3.87$. }
\label{Figsim}
\end{figure}

In fig.~\ref{Figsim}, we present the numerically obtained equilibrium profiles of $\phi(\bm x)$.
We examine the phase separation by varying  the composition $\phi_0$ and the $\chi$ values
as marked with the four crosses (A)--(D) in fig.~\ref{Figchis}(c).
As expected, we clearly see microphase separations in (A)--(C) and a macrophase separation in (D). 
In (A) and (B), hexagonal and stripe patterns are obtained, respectively. 
The characteristic lengths are approximately (A) $2.6$~nm and (B) $3.0$~nm within our parameters.
In fig.~\ref{Figsim}(C), on the other hand, there is no periodic pattern but ring-like aggregates are formed. 
For the parameter values in (C), a macrophase separated state is also a metastable state where the 
composition profile is similar to that in (D). 
We have calculated the grand potential $\Omega$ defined after eq.~(\ref{LipidEQ}) for both 
the microphase and macrophase separated states, and found that $\Omega$ for the microphase is 
smaller than that of the macrophase. 
This suggests that the transition between the micro and the macro phases is discontinuous. This is analogous to the first order unbinding transition in an amphiphile-water mixture, where abrupt swelling of lamellar phases takes place \cite{Schick}.
The full phase diagram including the lower temperature region should be further studied in the future.

\section{Conclusion} 

In this Letter, we have investigated the microphase formation in charged 
membranes, where the membrane can have non-zero net charge. 
Without assuming the DH condition, 
we obtained the solution eq.~(\ref{PkApp}) of the nonlinear PBE when the charge heterogeneity in 
the membrane is small. 
Using the solution, we calculated the static structure factor $S(k)$ of the membrane composition. 
We then discussed the microphase separation in a binary membrane composed of anionic and neutral 
lipids, for which the DH approximation is 
not justified except for a very small fraction of the charged lipid. 
Our theory reveals that the characteristic wave number $k^*$, at which $S(k)$ takes a maximum value, 
corresponds to the scale in the range from several to tens of nanometers except in the vicinity of the MEP. 
This explains why microphase separated structures have not been observed by optical microscopy 
measurements. 
We further predict a charge-induced MEP in the composition-temperature ($\phi_0, \chi$)-plane, 
which cannot be obtained within the DH approximation. 
The numerical simulation also shows that our model exhibits both the microphase and macrophase 
separations depending on the composition and/or the temperature.

We make further remarks. 
(i) For small scales corresponding to our predicted values of $k^*$, it would be preferable to use 
a microscopic density functional free energy for $F_{\rm m}$ rather than the mean field $f(\phi)$ and 
the gradient expansion form in eq.~(\ref{Fl}). 
Nevertheless, we believe that the present approximation provides reliable predictions
of the nano-domain formation.
(ii) The DH approximation is valid for a very small charge density $\sigma_0$ in the homogeneous 
state and/or in the weak segregation regime, as discussed in this Letter. 
However, even if $\sigma_0$ is very small,  it is not justified in the strong segregation regime 
where the charge density in each domain  becomes large. 
Without assuming the DH condition, Naydenov {\it et al.} discussed the strong segregation 
regime of a membrane that has no net charge, $\sigma_0=0$~\cite{Safran}. 
(iii) The solution eq.~(\ref{PkApp}) has further applications, such as the electrostatic contribution 
to the bending rigidity~\cite{Guttman}, and the charge regulation effect~\cite{Israelachivili} on a 
surface which has a spatially heterogeneous ionizable group distribution.

\acknowledgments

R.O. and S.K. thank C.\ Watanabe for informative discussion.
N.S. acknowledges support from the Grant-in-Aid for Young Scientist (B) (Grant No.\ 26800222) 
from the Japan Society for the Promotion of Science (JSPS) and the Grant-in-Aid for Scientific Research on Innovative Areas “{\it Molecular Robotics}” 
(Grant No.\ 15H00806) from the Ministry of Education, Culture, Sports, Science, and Technology of Japan (MEXT). 
S.K. acknowledges support from the Grant-in-Aid for Scientific Research on
Innovative Areas ``{\it Fluctuation and Structure}" (Grant No.\ 25103010) from the MEXT,
the Grant-in-Aid for Scientific Research (C) (Grant No.\ 24540439)
from the JSPS,
and the JSPS Core-to-Core Program ``{\it International Research Network
for Non-equilibrium Dynamics of Soft Matter}".

\end{document}